\documentclass[aps,prl,twocolumn,superscriptaddress,amsmath,amssymb,letterpaper]{revtex4}

\ifx\pdfoutput\@undefined\usepackage[usenames,dvips]{color}
\else\usepackage[usenames,dvipsnames]{color}
\usepackage{graphicx}
\usepackage{bm}
\usepackage{amssymb}
\usepackage{hyperref}
\usepackage{color}

\usepackage{amsmath}

\usepackage[T1]{fontenc}
\usepackage[latin1]{inputenc}

\newcommand{\be}{\begin{equation}}
\newcommand{\ee}{\end{equation}}

\newcommand{\bs}{\boldsymbol}

\begin{document}

\title{Edge Modes, Degeneracies, and Topological Numbers in Non-Hermitian Systems}

\author{Daniel Leykam}
\affiliation{Division of Physics and Applied Physics, School of Physical and Mathematical Sciences, Nanyang Technological University,
Singapore 637371, Singapore}

\author{Konstantin Y. Bliokh}
\affiliation{CEMS, RIKEN, Wako-shi, Saitama 351-0198, Japan}
\affiliation{Nonlinear Physics Centre, RSPE, The Australian National University,
Canberra, ACT 0200, Australia}

\author{Chunli Huang}
\affiliation{Division of Physics and Applied Physics, School of Physical and Mathematical Sciences, Nanyang Technological University,
Singapore 637371, Singapore}
\affiliation{Department of Physics, National Tsing Hua University, Hsinchu 30013, Taiwan}

\author{Y. D. Chong}
\affiliation{Division of Physics and Applied Physics, School of Physical and Mathematical Sciences, Nanyang Technological University,
Singapore 637371, Singapore}

\affiliation{Centre for Disruptive Photonic Technologies, Nanyang Technological University, Singapore, 637371, Singapore}

\author{Franco Nori}
\affiliation{CEMS, RIKEN, Wako-shi, Saitama 351-0198, Japan}
\affiliation{Physics Department, University of Michigan, Ann Arbor, Michigan 48109-1040, USA}

\begin{abstract}
We analyze chiral topological edge modes in a non-Hermitian variant of the 2D Dirac equation. Such modes appear at interfaces between media with different ``masses'' and/or signs of the ``non-Hermitian charge''. The existence of these edge modes is intimately related to {\it exceptional points} of the bulk Hamiltonians, i.e., degeneracies in the bulk spectra of the media. We find that the topological edge modes can be divided into {\it three families} (``Hermitian-like'', ``non-Hermitian'', and ``mixed''); these are characterized by {\it two winding numbers}, describing two distinct kinds of half-integer charges carried by the exceptional points. We show that all the above types of topological edge modes can be realized in honeycomb lattices of ring resonators with asymmetric or gain/loss couplings.
\end{abstract}

\vspace*{-0.1cm}

\maketitle

\textit{Introduction.---}
There is presently enormous interest in two groups of fundamental physical phenomena: (i) topological edge modes in quantum Hall fluids and topological insulators~\cite{Thouless,TI_reviews,Lu2014}, which are Hermitian, and (ii) novel effects in non-Hermitian wave systems (including $\mathcal{PT}$-symmetric systems)~\cite{non-Hermitian_review,Bender2007,Cao2015}. Both types of phenomena have been studied in the context of quantum as well as classical waves, and both are deeply tied to the geometrical features of spectral degeneracies. 
In the Hermitian case, the common degeneracies are Dirac points: linear band-crossings (generically, in a 3D parameter space), which separate distinct topological phases and mark the birth or destruction of topological edge modes~\cite{dirac_review_1,dirac_review_2}. Non-Hermitian systems, however, exhibit a distinct class of spectral degeneracies known as {\it exceptional points} (EPs), which are branch points in a 2D parameter space where the Hamiltonian becomes non-diagonalizable~\cite{non-Hermitian_review,Berry2004,Heiss2012}.

In Hermitian systems, the bulk-edge correspondence relations that give rise to topological edge modes are typically based upon the Berry connection, which in turn relies on the eigenvector orthogonality granted by Hermiticity~\cite{Thouless,TI_reviews,Lu2014}. Is there a generalization of the bulk-edge correspondence to non-Hermitian systems~\cite{Rudner2009}?  When sufficiently weak non-Hermiticity (e.g.~loss) is introduced to topological insulator models, the edge modes can retain some of their original characteristics~\cite{Esaki,hu2011}.  On the other hand, certain non-Hermitian models with chiral symmetry can support anomalous edge modes that have no clear relationship to Hermitian topological edge modes~\cite{Schomerus,Malzard2015,Lee}. These modes are embedded within a complex gapless band structure, and appear in the vicinity of EPs; however, it is not known whether they can be related to model-independent bulk topological invariants similar to those in Hermitian systems.

This paper aims to shed light on the nature of topological edge modes in non-Hermitian quantum systems. In contrast to previous studies based on lattice models~\cite{Rudner2009,Esaki,hu2011,Schomerus,Malzard2015,Lee,Diehl2011,Zeuner2015,Yuce2015,SanJose2016,Gonzales2016,Rudner2016}, we focus on a non-Hermitian {\it continuum} model. This is motivated by the fact that, in the Hermitian case, many model-independent features of topological edge modes can be understood in terms of the generic properties of continuum models, such as the Dirac equation in various dimensions \cite{dirac_review_1,JR,Schnyder,Shen2011}. For example, zero-energy Jackiw-Rebbi end modes of the 1D Dirac equation \cite{JR,Shen2011} unperpin end modes of the SSH lattice model \cite{SSH,Kitaev}.

Our continuum model consists of a 2D non-Hermitian Hamiltonian that is linear in both $k_x$ and $k_y$ and possesses a tunable mass parameter $m$, similar to the 2D Dirac equation.  
The bulk band structure is complex and possesses a pair of EPs (branch points). Along interfaces between media with different ``masses'' and/or signs of the non-Hermiticity, we find that there exist {\it zero-energy chiral edge modes}. Remarkably, the appearance of these edge modes and their regions of existence are fully determined by the EPs in the bulk spectra of the media.  We show that these modes can be classified as ``Hermitian-like'', ``non-Hermitian'', and ``mixed'', using {\it two topological numbers}. The first number is related to the chirality of the eigenstates (i.e., the sign of the Berry curvature), while the second one characterizes the chirality of the EP~\cite{Heiss2001,Dembowski2003,Peng2016,Berry2004,Heiss2012}. The ``non-Hermitian'' and ``mixed'' edge modes resemble the ``anomalous'' edge modes found in Ref.~\cite{Lee}. Moreover, we are able to enumerate the zero-energy edge modes by using an {\it index theorem}, a variant of the Aharonov-Casher theorem for the 2D Dirac equation in a vector potential \cite{AharonovCasher}. Finally, we show that a lattice counterpart of this continuum model, including the anomalous edge modes, can be realized using honeycomb-like arrays of ring resonators with non-Hermitian couplings~\cite{longhi2015,supplemental,doubling}.

\textit{Non-Hermitian Hamiltonian.---}
%
Our model is based on the following non-Hermitian Hamiltonian, defined on a 2D momentum space ${\bf k} = (k_x,k_y)$:
\begin{align}
  \begin{aligned}
    \hat{H} &= \begin{pmatrix} k_x - i s k_y & m \\ m & -k_x + i s k_y \end{pmatrix}  
    \\ &\equiv
\left( k_x - i s k_y  \right) \hat{\sigma}_z + m \hat{\sigma}_x \equiv \bf{B} \cdot \hat{\bm \sigma}.
  \end{aligned}
  \label{eq:hamiltonian}
\end{align}
Here, $\hat{\bm \sigma} = (\hat{\sigma}_z, \hat{\sigma}_x, \hat{\sigma}_y)$ denotes the vector of Pauli matrices (permuted cyclically for later convenience), and ${\bf B} = (B_x,B_y,0)$ is an effective complex ``magnetic field'', which will be used in the subsequent Berry-phase analysis. The Hamiltonian $\hat{H}$ contains three continuously-tunable real parameters: the momenta $k_x$ and $k_y$, and $m$ (assumed real), which mixes the two spinor components, and which we call ``mass'' for convenience. The parameter $s=\pm 1$, which we will regard as a ``non-Hermitian charge'', determines the sign of the imaginary part, such that $[H(s)]^\dagger = H(-s)$.

\begin{figure}
\centering
\includegraphics[width=\columnwidth]{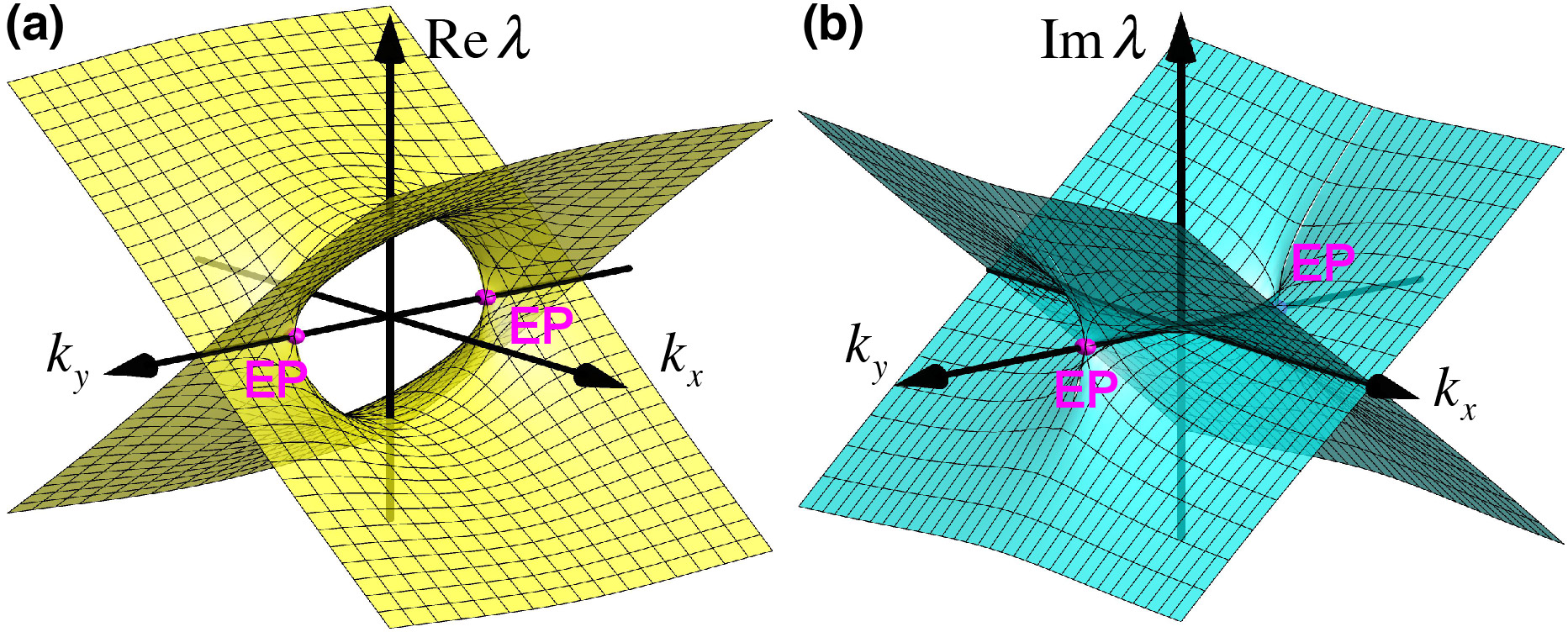}
\caption{Real and imaginary parts of the complex spectrum (\ref{spectrum}) of the Hamiltonian (\ref{eq:hamiltonian}) with exceptional points at ${\bf k}_{\rm EP}^\pm = (0,\pm |m|)$.}
\label{fig:spectrum}
\end{figure}

The Hamiltonian (\ref{eq:hamiltonian}) involves only two Pauli matrices, and has the chiral symmetry $\{\hat{H},\hat{\sigma}_y\} = 0$. It is also $\mathcal{PT}$ symmetric (where $\mathcal{T}$ involves complex conjugation and momenta reversal, while $\mathcal{P}$ is the reflection $x \rightarrow -x$), and can hence have real eigenvalues~\cite{Bender2007,Cao2015}.  The eigenvalues of $\hat{H}$ are
\be 
\lambda^{\pm} = \pm \sqrt{{\bf B} \cdot {\bf B}} = \pm \sqrt{m^2 + (k_x - i s k_y)^2}\,,
\label{spectrum}
\ee
and its (non-normalized) eigenvectors are
\begin{equation}
  \psi^{\pm} = 
\begin{pmatrix} 1 \\ B_y/(B_x+\lambda^{\pm}) \end{pmatrix}~.
\label{eq:spectrum}
\end{equation}
The complex spectrum (\ref{spectrum}) is shown in Fig.~\ref{fig:spectrum}.  Along the $k_y$ axis, the real part of the spectrum, $\mathrm{Re}(\lambda)$\textcolor{red}{,} is gapped for $-|m| < k_y < |m|$, and ungapped for $|k_y| > |m|$.  There are two EPs at ${\bf k}_{\rm EP} = (0,\pm |m|)$, separating the ``gapped'' and ``ungapped'' $k_y$-domains.

Unlike Hermitian degeneracies, EPs involve the coalescence of eigenvectors, not just eigenvalues $\lambda^{\pm}({\bf k}_{\rm EP})=0$. $\hat{H}({\bf k}_{\rm EP})$ is defective and has a single chiral eigenmode (an eigenvector of $\hat{\sigma}_y$):
\begin{equation}
  \psi({\bf k}_{\rm EP})
  = \begin{pmatrix} 1 \\ i \chi_{\rm EP} \end{pmatrix}~, 
\label{eq:charge_1}
\end{equation}
where $\chi_{\rm EP} = \pm {\rm sgn}(sm)$ is the chirality of the EP~\cite{Heiss2001,Dembowski2003,Peng2016,Berry2004,Heiss2012}.
%

\textit{Chiral edge modes.---}
%
We translate Eq.~(\ref{eq:hamiltonian}) to a Schr\"odinger wave equation by taking $\hat{\bf k} = - i \nabla$ and allowing the mass $m$ and/or non-Hermitian charge $s$ to vary with position (though we still assume that $s$ only takes the values $\pm 1$):
\begin{equation}
  \hat{H} =
  \big[ -i\partial_x - s(x,y)\, \partial_y  \,\big] \hat{\sigma}_z
  + m(x,y) \hat{\sigma}_x.
  \label{wave_eq}
\end{equation}

Consider an interface between two uniform media with different $m$ and/or $s$. 
For now, let the interface be along the line $x = 0$, such that $m=m_1,~s=s_1$ for $x<0$ (medium 1), and $m=m_2,~s=s_2$ for $x>0$ (medium 2).  
We seek edge modes that propagate along $y$ and are normalizable along $x$:
\be 
\psi_{\rm edge} = \left(\begin{array}{c} \alpha \\ \beta \end{array} \right) \begin{cases} e^{i k y + \kappa_1 x}, & \mathrm{Re}(\kappa_1)<0,~~x > 0 \\ e^{i k y + \kappa_2 x}, & \mathrm{Re}(\kappa_2)>0,~~x < 0. \end{cases}
\label{edge_mode}
\ee
Substituting Eq.~(\ref{edge_mode}) into Eq.~(\ref{wave_eq}), we find the {\it zero-energy edge modes}, $\lambda_{\rm edge} = 0$, which exist when the following real equations are satisfied:
\be 
\quad -\kappa_1 =  s_1 k \pm m_1, \quad -\kappa_2 = s_2 k \pm m_2. \label{eq:edge_modes}
\ee
For $\kappa_1<0$ and $\kappa_2>0$, there can be zero, one, or two solutions to Eq.~(\ref{eq:edge_modes}) for each $k$.  The number of solutions also depends on $m_{1,2}$ and $s_{1,2}$.  Like the eigenmodes at the EPs of the bulk system, these edge modes are {\it chiral}, satisfying $\beta/\alpha = \pm i$.  Similar to Eq.~(\ref{eq:charge_1}), we define the mode chirality as $\chi_{\rm edge} = {\rm Im} \left(\beta/\alpha\right)_{\rm edge}$.

We first examine the two simplest cases:

(A) The media have equal charges, $s_1=s_2=s$, and opposite masses, $m_1=-m_2=m$. In this case, there is \textit{one} zero-energy edge mode for each $k \in (-|m|, |m|)$, and no edge modes for all other $k$. This $k$-range corresponds to the $k_y$-domain with the {\it gapped} bulk spectra $\mathrm{Re}(\lambda^{\pm})$ between the two EPs (Fig.~\ref{fig:spectrum}). This domain includes $k=0$, which is the {\it Hermitian} limit where Eq.~(\ref{wave_eq}) reduces to the {\it Jackiw-Rebbi} model for 1D Dirac modes \cite{JR,Shen2011}. Thus, this is a family of non-Hermitian edge modes that are continuable from the Hermitian Jackiw-Rebbi edge modes. The mode chirality is $\chi_{\rm edge} = {\rm sgn} (m)$, independent of $s$.

(B) The media have equal ``masses'', $m_1=m_2=m$, but opposite ``charges'', $s_1=-s_2 \equiv s$. In this case, there are {\it two} edge modes in the domain $k \in {\rm sgn}(s)(|m|, \infty)$. This corresponds to one of the $k_y$ domains with the {\it ungapped} $\mathrm{Re}(\lambda^{\pm})$ bulk spectra. The two edge modes have opposite chiralities, $\chi_{\rm edge} = \pm 1$, and are independent of $m$. Unlike case (A), these modes are essentially {\it non-Hermitian}. First, they are asymmetric in $k$, and do not exist in the Hermitian limit $k=0$. Second, the modes are {\it defective}: the corresponding left eigenvectors (right eigenvectors of $\hat{H}(-k) = \hat{H}^{\dagger}(k)$) do not exist.

\begin{figure}
\centering
\includegraphics[width=\columnwidth]{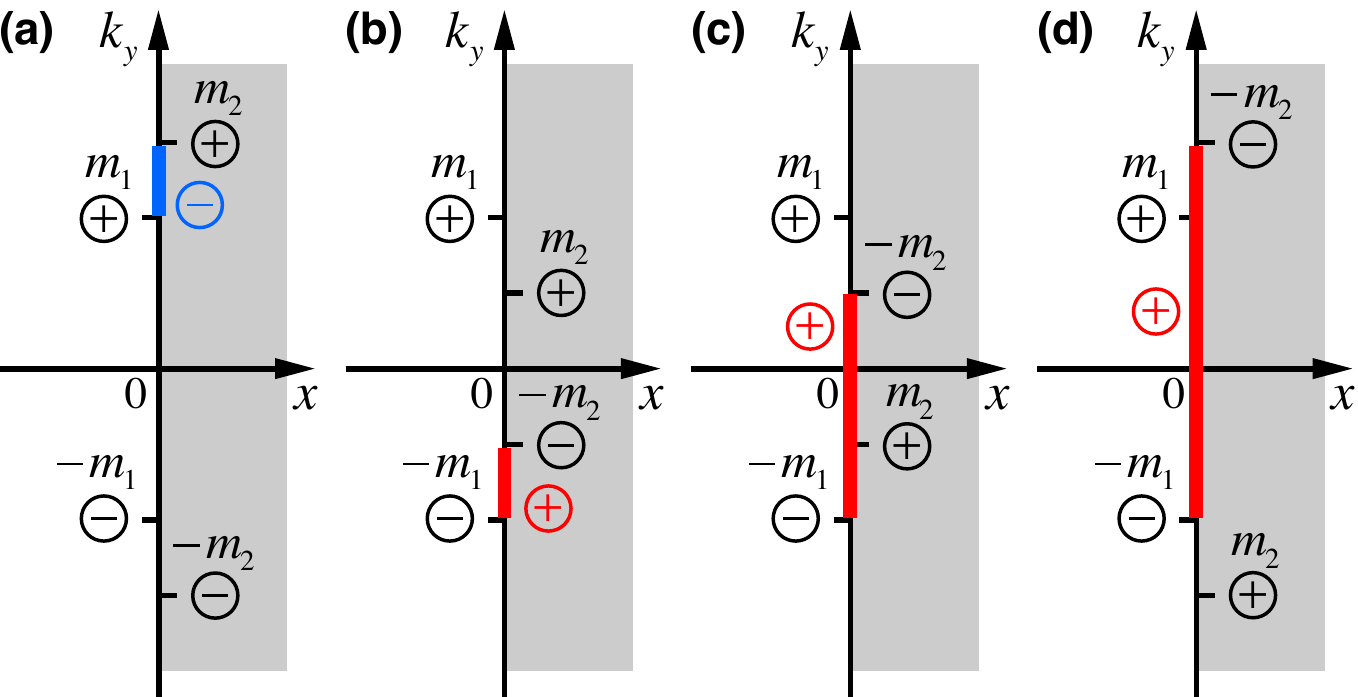}
\caption{Schematic diagrams indicating the zero-energy chiral edge modes (\ref{edge_mode}) and (\ref{eq:edge_modes}) at the interface $x=0$ between two media with different ``masses'' $m_1>0$ and $m_2$,  and the same ``non-Hermitian charges'' $s_1=s_2=1$.
Exceptional points ${\bf k}_{\rm EP}$ are indicated for the two media, with chiralities $\chi_{\rm EP}$ marked by the black ``$+$'' and ``$-$'' signs. The edge modes with positive and negative chiralities $\chi_{\rm edge}$ are marked by red and blue colors.}
\label{fig:2}
\end{figure}

When $|m_{1}| \neq |m_2|$, the situation is more complicated.  Fig.~\ref{fig:2} shows the edge modes for varying $m_2$, with $s_1 = s_2 = 1$ and $m_1 > 0$.  For $m_2 > m_1$, there is one edge mode for each $k \in (m_1, m_2)$, as shown in Fig.~\ref{fig:2}(a).  For $m_2 < m_1$, there is one edge mode for each $k \in (-m_1, -m_2)$, as shown in Fig.~\ref{fig:2}(b)--(d); this includes the special case (A) discussed above.  For certain values of $k$, ${\rm Re}(\lambda^{\pm})$ is gapped in one medium and ungapped in the other medium. We call such zones and the corresponding edge modes {\it ``mixed''}. When $m_2 > 0$, there are only positive or only negative values of $k$, as shown in Fig.~\ref{fig:2}(a),(b).  In Fig.~\ref{fig:2}, we also indicate the chiralities of the EPs in the two media, $\chi_{\rm EP}$, and the chiralities of the edge modes, $\chi_{\rm edge}$. Notably, the edge modes always connect a pair of EPs with the same chirality, while the modes themselves have the opposite chirality.

\begin{figure}
\centering
\includegraphics[width=0.7\columnwidth]{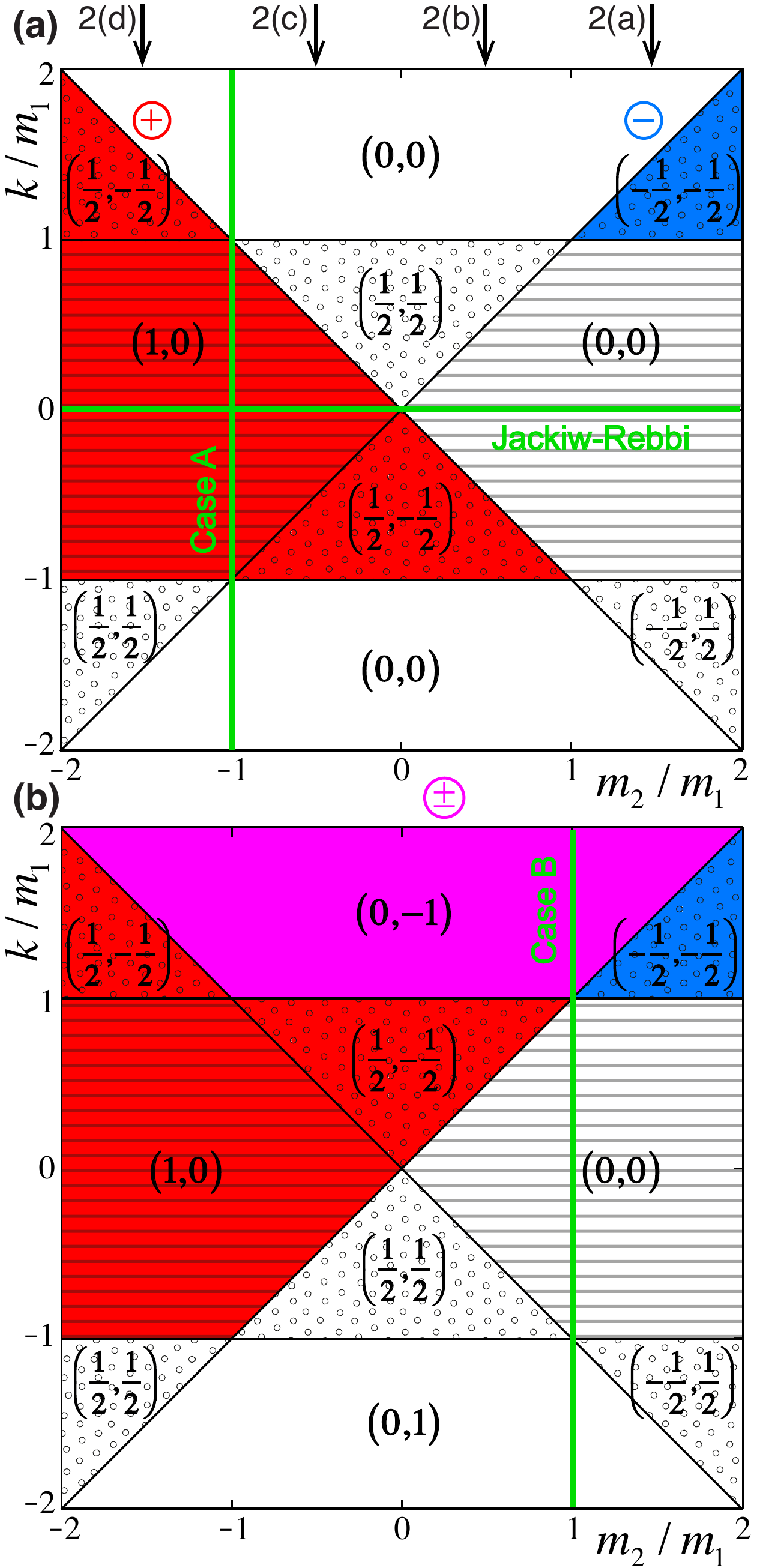}
\caption{Phase diagrams for the chiral edge modes (\ref{edge_mode}) and (\ref{eq:edge_modes}) at the interface between two media with (a) equal ``non-Hermitian charges'' $s_1=s_2=1$, and (b) opposite ``charges'' $s_1=-s_2=1$.  The ``mass'' in medium 1 is fixed as $m_1>0$.  The arrows above (a) indicate the cases shown in Fig.~\ref{fig:2}.  The numbers in parentheses indicate the differences of winding numbers (\ref{eq:w1}) and (\ref{eq:w2}) between the two media: $(\Delta w_1,\Delta w_2)$.
Striped, empty, and dotted zones indicate gapped, ungapped, and mixed cases of the bulk spectra ${\rm Re}(\lambda^{\pm})$ in the two media.}
\label{fig:3}
\end{figure}

We summarize the conditions under which zero-energy edge modes exist using the phase diagrams in Fig.~\ref{fig:3}. Here, we fix $m_1>0$ and $s_1=1$, and use $k/m_1$ and $m_2/m_1$ as plot axes. The red (blue) regions show where there is a single edge mode with $\chi_{\rm edge}=+1$ ($\chi_{\rm edge}=-1$). Figure~\ref{fig:3}(a) shows the case where the two media have equal non-Hermitian charge $s$, with the special case (A) lying on the $m_2/m_1 = -1$ line and the Jackiw-Rebbi model \cite{JR,Shen2011} lying on the $k=0$ line. Figure~\ref{fig:3}(b) shows the opposite-charge case; it also contains a purple region indicating two edge modes with $\chi_{\rm edge} = \pm 1$, which includes the case (B) on the line $m_2/m_1 = 1$.

We will now show that these phase diagram features---i.e., the number of zero-energy edge modes and under what conditions they appear---can be understood from  the topological properties of Eq.~(\ref{eq:hamiltonian}).

\textit{Winding numbers.---}
%
Since one family of edge modes can be continued to Jackiw-Rebbi modes \cite{JR,Shen2011}, and the termination points of the edge modes are EPs of the bulk spectrum, we can guess that the edge modes can be  characterized by bulk topological invariants \cite{Thouless,TI_reviews,Lu2014}.  Along the interface, the conserved $k_y$ plays the role of a tunable parameter for calculating a 1D winding number. However, it turns out that \textit{two} winding numbers are needed to fully describe the edge modes in the non-Hermitian case.

Previous researchers \cite{mailybaev2005,Rudner2009,Esaki,Lee} have focused on the winding numbers of the eigenvectors $\psi^{\pm}$. However, we emphasize that encircling an EP (branch point) swaps the bands, so that {\it two} loops in parameter space are required to return to the original state (with a $\pi$ geometric phase gained) \cite{non-Hermitian_review,Berry2004,Heiss2012,mailybaev2005,Lee,demowski2004,gao2015}. Hence, there is no globally smooth way to define two distinct bands for Eq.~\eqref{eq:spectrum}.


One way to resolve this band-labelling ambiguity is to consider winding numbers associated with the complex ``magnetic field'' $\bs{B}$ defined in Eq.~(\ref{eq:hamiltonian}), which has no discontinuities.  We take a spherical-like representation $\bs{B} = B (\sin \theta \cos \phi, \sin \theta \sin \phi, \cos \theta)$, where both the ``magnitude'' $B = \lambda^{+}$ and the ``angles'' $(\theta,\phi)$ are complex~\cite{garrison1988}.  The chiral symmetry of $\hat{H}$ constrains $\bs{B}$ to the plane $\theta=\pi/2$, so only $B$ and $\phi$ vary with $\mathbf{k}$.

We now introduce the winding number
\be 
w_1 = \frac{1}{2\pi} \int_{k_x=-\infty}^{k_x=+\infty} \nabla_k \phi \cdot d\mathbf{k},
\label{eq:w_1}
\ee
where the integral is taken along a $k_x$-line with fixed $k_y$.  This winding number orginates from a non-Hermitian generalization of the Berry phase, describing the effects of varying ``{\it direction}'' of $\bs{B}$~\cite{garrison1988}. It is equivalent to the winding numbers used in Refs.~\cite{mailybaev2005,Rudner2009,Esaki}.  Applying Eq.~\eqref{eq:w_1} to Eq.~\eqref{eq:hamiltonian}, we find \cite{supplemental}
\begin{align} 
w_1 
&= \begin{cases} -\frac{1}{2} \mathrm{sgn}(m), & \mathrm{for}\;|k_y| < |m|
\\ 0, & \mathrm{for}\;|k_y| > |m|. \end{cases} 
\label{eq:w1}
\end{align}
This explains the edge modes in case (A), corresponding to the $m_2/m_1=-1$ line in Fig.~\ref{fig:3}(a). The difference in the topological numbers of the two media is $\Delta w_1 = w_1(m_2) - w_1(m_1) = {\rm sgn} (m_1)$; accordingly, we observe a single edge mode of chirality $\chi_{\rm edge} = {\rm sgn}(m_1)$.

For other parameter choices, $\Delta w_1$ can be fractional.  For Fig.~\ref{fig:2}(a) we find $\Delta w_1 = -1/2$, and for Fig.~\ref{fig:2}(b), $\Delta w_1 = 1/2$.  Edge modes in these cases resemble the ``anomalous'' edge modes found in Ref.~\cite{Lee}. Clearly, $w_1$ alone is insufficient to characterize these modes, which are asymmetric in $k$.

To classify the anomalous edge modes, we introduce a second winding number using the complex ``{\it magnitude}'' of ${\bf B}$: $B = \lambda^{+}$. Near each EP, the eigenvalues form ``half-vortices'': $\lambda^{\pm} \propto \pm \sqrt{|{\bf k}-{\bf k}_{\rm EP}|} \exp \left[ i s {\rm Arg}({\bf k}-{\bf k}_{\rm EP})/ 2 \right]$, where $s/2$ is the vortex topological charge.  We define
\be 
w_2 = \frac{1}{2\pi} \int_{k_x=-\infty}^{k_x=+\infty} \nabla_k \mathrm{Arg}(\lambda^{+}) \cdot d\mathbf{k},
\ee
where the integral is again taken similarly to Eq.~\eqref{eq:w_1}.  For the spectrum \eqref{spectrum}, we find \cite{supplemental}
\be 
w_2 = \begin{cases} 0, & |s k_y| < |m| \\
\frac{1}{2} \mathrm{sgn}(s k_y), & |s k_y| > |m|.
\end{cases}
\label{eq:w2}
\ee
This winding number has the required asymmetry in $k_y$.  Whenever $w_2 \ne 0$, there are branch cuts in $\lambda^{\pm}$, and $\hat{H}$ cannot be continuously deformed into a gapped Hermitian system. Unlike $w_1$, which is a generalization of the Berry phase, the $w_2$ winding number is specific to non-Hermitian systems and has no direct Hermitian counterpart.

Using $\Delta w_1$ and $\Delta w_2$, we can completely characterize the edge modes shown in Fig.~\ref{fig:3}.  First, for $w_2 = 0$, the existence of ``Hermitian-like'' edge modes (and their chirality) is determined by $\Delta w_1$. Second, for $\Delta w_2 \ne 0$, the number of anomalous (``non-Hermitian'' and ``mixed'') edge modes is $2|\Delta w_2|$, while $\mathrm{sgn}(\Delta w_2)$ determines whether they are localized to the left or right edge of medium 1. In Fig.~\ref{fig:3}, the anomalous non-Hermitian edge modes only exist on the right edge when $\Delta w_2 < 0$.  In particular, the purple region in Fig.~\ref{fig:3}(b) corresponds to $\Delta w_2 = -1$, and accordingly there are two anomalous edge modes with opposite chiralities ($\Delta w_1 = 0$), and both are defective. Thus, the winding numbers $w_{1,2}$ provide the bulk-edge correspondence for the non-Hermitian Hamiltonian (\ref{wave_eq}) and describe topological properties of the edge modes Fig.~\ref{fig:3}.

Since $w_1$ and $w_2$ only change when $k_y$ crosses an EP, we can identify the ``topological charges'' of the individual EPs as $(q_1,q_2)=\frac{1}{2} (\pm |m|, s)$. There are {\it four} inequivalent non-Hermitian degeneracies, in contrast to the {\it two} inequivalent Hermitian degeneracies. This is a consequence of the richer morphologies of complex vector fields that parametrize non-Hermitian Hamiltonians~\cite{dennis_review}. 

\textit{Index theorem.---}
Another way to analyze the zero-energy modes (zero modes) of the non-Hermitian Hamiltonian (\ref{wave_eq}) is to consider the Hermitian Hamiltonian
\begin{equation}
  \hat{\mathcal{H}} = \hat{H}^\dagger \hat{H}.
\end{equation}
Zero modes of $\hat{H}$ are also zero modes of $\hat{\mathcal{H}}$, and vice versa.  When $s=\pm 1$ is a constant, we find that
\begin{equation}
  \hat{\mathcal{H}}
  =\left|-i\nabla - \hat{\sigma}_y s\bm{\mathcal A}(x,y)\right|^{2} + \hat{\sigma}_y \mathcal{B}(x,y),
\end{equation}
where $\mathcal{B}(x,y)=\partial_{x}\mathcal{A}_{y}-\partial_{y}\mathcal{A}_{x}$ and $\bm{\mathcal A} = \left(0, m\right)$. This is a Pauli-type Hamiltonian for a nonrelativistic particle in a matrix-valued vector potential \cite{Jackiw}.

The normalizable zero modes of $\hat{\mathcal{H}}$ can now be counted by an ``index theorem'' argument \cite{AharonovCasher}.  The result is that there are $N =\lfloor |\Phi | / 2\pi \rfloor$ such modes, where $\Phi$ is the total flux of $B$. This holds for {\it arbitrary} complex analytic mass fields $m(x,y)$.  For the previously-considered special case of two media with a straight interface, there is a flux of $(m_2 - m_1)$ per unit length along the domain wall, implying that the zero modes occupy a $k$-range of $\Delta k = m_2 - m_1$, in precise agreement with Fig.~\ref{fig:2} and Fig.~\ref{fig:3}(a) (see details in \cite{supplemental}).

\textit{Discussion.---}
We have analyzed a 2D non-Hermitian continuum model that exhibits different types of zero-energy edge modes, which can be classified using two half-integer-valued winding numbers calculated from the complex bulk band structure. These are inherently associated with topological properties of bulk eigenmodes and non-Hermitian degeneracies (EPs) in the band structure. One family of edge modes includes the well known (Hermitian) Jackiw-Rebbi zero modes \cite{JR,Shen2011}.  However, the classification also contains essentially non-Hermitian edge modes that cannot be continued into Jackiw-Rebbi-type edge modes; these seem to be continuum counterparts of the ``anomalous'' edge modes recently encountered in certain 1D non-Hermitian lattice models \cite{Schomerus,Malzard2015,Lee}.

The three families of non-Hermitian topological edge modes can be realized in a non-Hermitian 2D photonic resonator lattice \cite{Cao2015,Schomerus,Malzard2015,Szameit,Ramezani,Liang2013,Chong2015}, with non-Hermicity introduced through either asymmetric scattering between clockwise and anticlockwise modes~\cite{non-Hermitian_review,Bender2007,Cao2015,Schomerus,Malzard2015}, or amplifying/lossy inter-resonator coupling \cite{Ramezani,Liang2013,Chong2015}.  In the Supplemental Materials \cite{supplemental}, we show that lattice and interface orientations can be chosen to yield different values of $(w_1,w_2)$ and, correspondingly, different families of zero-energy edge modes~\cite{doubling}.

We have focused on the case of two uniform media separated by the line $x = 0$.  For other orientations of a straight interface, we obtain similar phase diagrams, taking $k = \bs{k}\cdot \hat{\bs{y}}$ where $\bs{k}$ is the wavevector parallel to the interface. The index-theorem derivation of the number of normalizable zero modes is even more general, and applies to arbitrary analytic mass fields. The above features, and comparisons with previously-known examples, suggest that the variety of chiral edge modes and topological numbers found in this work may be generic to a wide class of non-Hermitian wave systems.

\vspace*{0.2cm}
\begin{acknowledgments}
The idea of this research was conceived by K.Y.B., C.Y.D., and D.L. at the workshop ``Topological States of Light and Beyond'' in the Center for Theoretical Physics of Complex Systems, Institute for Basic Science, Daejeon, South Korea. We are grateful to A. Miroshnichenko, A. B. Khanikaev, and S. Flach for their hospitality and inspiring discussions. This research was supported by the Singapore National Research Foundation (grant NRFF2012-02), the Singapore MOE Academic Research Fund Tier 2 (grant MOE2015-T2-2-008), the RIKEN iTHES Project, the MURI Center for Dynamic Magneto-Optics via the AFOSR (grant FA9550-14-1-0040), Grant-in-Aid for Scientific Research (A), the John Templeton Foundation, and the Australian Research Council.
\end{acknowledgments}

\clearpage

\renewcommand{\theequation}{S\arabic{equation}}
\makeatother
\setcounter{equation}{0}
\makeatletter
\renewcommand{\thefigure}{S\@arabic\c@figure}
\makeatother
\setcounter{figure}{0}

\section{Supplementary Online Materials}

\section{I. Calculation of winding numbers}

We first illustrate the appearance of two winding numbers, Eqs.~(8)--(11) in the main text, from the properties of the complex effective ``magnetic field'' ${\bf B}$, Eq.~(1). This field lies in the $(x,y)$-plane, and can be characterized by its {\it magnitude} and azimuthal {\it direction} angle: ${\bf B} = B (\cos\phi, \sin\phi, 0)$.

First, the complex direction angle is given by
\begin{equation}
\phi =  \arctan \left(\frac{B_y}{B_x}\right) = \arctan \left(\frac{m}{k_x - i s k_y}\right).
\end{equation}
The distribution of its real part, ${\rm Re}\,\phi ({\bf k})$, is shown in Fig.~S1(a). Integrating its ${k_x}$-gradient along the countour shown in Fig.~S1(a) yields the first winding number $w_1(k_y)$, Eqs.~(8) and (9), shown in Fig.~S1(c). Note that only smooth gradients of ${\rm Re}\,\phi(k_x)$ contribute to the integral, but not $\pi$ jumps. Furthermore, the imaginary part of the angle $\phi$ does not contribute to the integrals (8) and (9). Therefore, the winding number $w_1$ is an extention of the Hermitian winding number based on the Berry phase and direction of the ${\bf B}$-field \cite{Rudner2009,Esaki}. 

Second, the magnitude of the ${\bf B}$-field is also {\it complex} in the non-Hermitian case. It is equal to the eigenvalue $\lambda^+$: 
\begin{equation}
B =  \sqrt{B_x^2 + B_y^2} = \sqrt{(k_x - i s k_y)^2 + m^2}=\lambda^+.
\end{equation}
The complex character of this quantity is characterized by its phase ${\rm Arg}\,\lambda^+$. The distribution ${\rm Arg}\,\lambda^+ (\bf k)$ is depicted in Fig.~S1(b). Integrating its $k_x$-gradient along the same contour yields the second winding number $w_2 (k_y)$, Eqs.~(10) and (11), shown in Fig.~S1(d). Since the nontrivial phase behavior of  eigenvalues is a purely non-Hermitian feature, the $w_2$ number characterizes ``non-Hermitian'' and ``mixed'' anomalous edge modes \cite{Schomerus,Lee}.

Importatly, the above description of the winding numbers, based on the properties of the complex ${\bf B}$-field, is universal and can be applied to other non-Hermitian two-level systems. Below we show this for two examples of 2D lattice systems.

\begin{figure}
\includegraphics[width=\columnwidth]{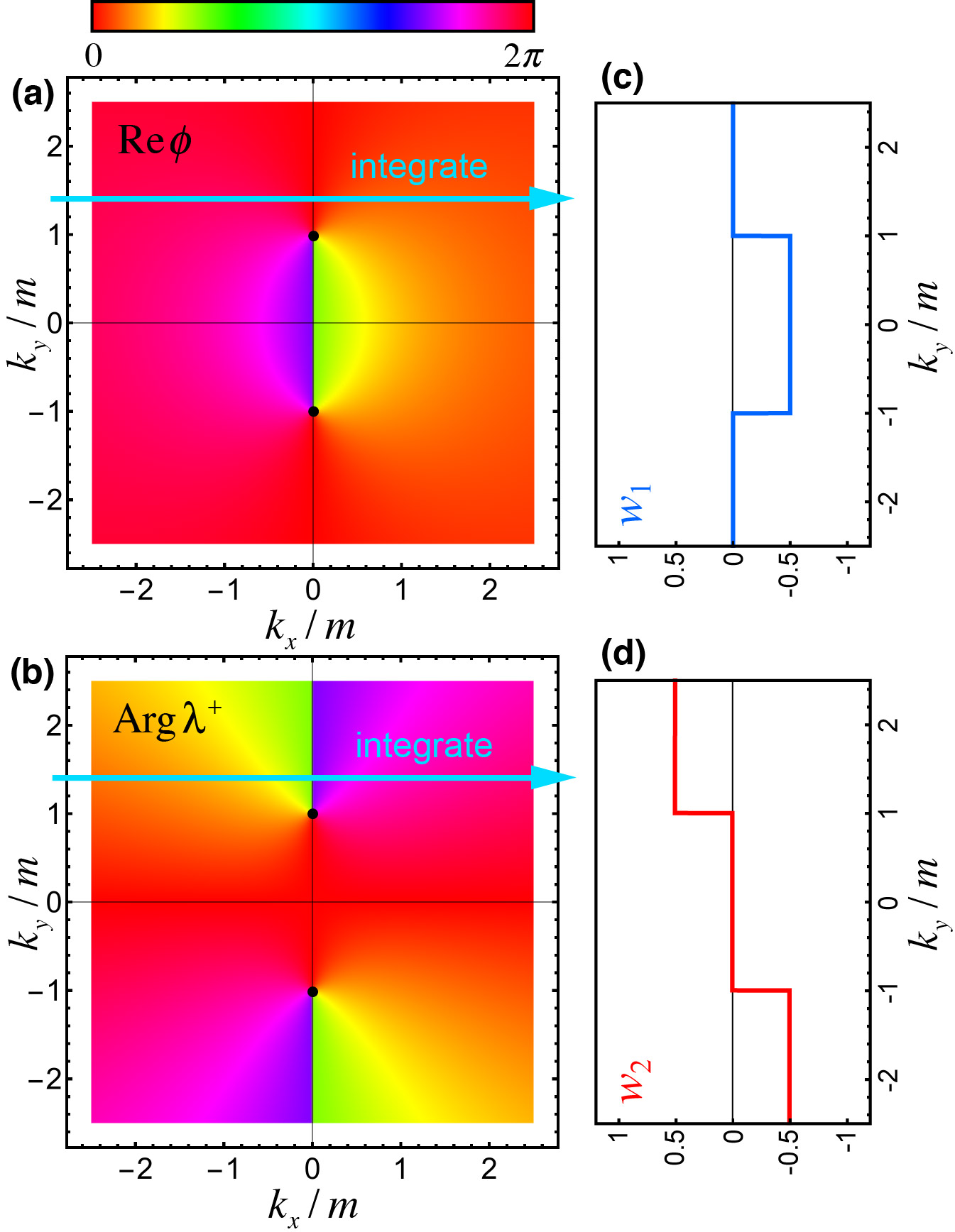}
\caption{Color-coded ${\bf k}$-distributions of (a) the real part of the azimuthal angle $\phi$ of the ${\bf B}$-field and (b) the phase of the complex field magnitude $B=\lambda^+$. Black dots indicate the EPs. Integration of the $k_x$-gradients of these angles along the $k_y$-dependent cyan contour yields the two winding numbers $w_1 (k_y)$ and $w_2 (k_y)$, Eqs.~(8)--(11), shown in panels (c) and (d).}
\label{fig:honeycomb_cell}
\end{figure}

\section{II. Tight-binding models}

One important distinction between the continuum model studied in the main text and lattice models is the periodic boundary conditions imposed by the 2D Brillouin zone in the latter. To satisfy periodic boundary conditions, any branch cut in the eigenvalues must terminate at an inequivalent EP, such that $\sum q_2 = 0$. Therefore, regularizing Eq.~(1) of the main text to a lattice will double the EP pairs, with the partners having opposite charges. This is analogous to the doubling of Dirac points in graphene-like systems implied by the Nielsen-Ninomiya theorem. Nevertheless, for most orientations of a lattice edge or domain wall, the doubled partners are decoupled. Nonzero winding numbers and edge modes will still occur within the finite range of momenta between the doubled partners~\cite{delplace2011}. 

Here we show how to achieve fractional winding numbers in ring resonator lattices. We consider anisotropic honeycomb lattices with inter-resonator coupling mediated by nonresonant link rings. In the Hermitian limit, the honeycomb lattice hosts Dirac point degeneracies characterized by integer winding numbers. Introducing non-Hermitian couplings by inserting gain or loss into the link rings splits the Dirac points into pairs of EPs, generating fractional winding numbers. We will consider two types of non-Hermitian couplings: balanced gain and loss within each link ring, which is described by a real asymmetric Hamiltonian~\cite{longhi2015}, and link rings with pure gain or loss, generating a complex symmetric Hamiltonian~\cite{longhi2016}. These two cases result in different trajectories of the EPs in momentum space as a function of the non-Hermiticity strength $\gamma$. 

\subsection{A. Model 1}

We begin by considering a non-Hermitian coupling generated by inserting balanced gain and loss into one of the three link rings, shown in Fig.~\ref{fig:honeycomb_cell}. For the clockwise-circulating resonator modes, intracell hopping from the ``A'' sublattice to the ``B'' sublattice occurs via the lossy half of the link resonator. Conversely, hopping from ``B'' to ``A'' occurs through the half of the link resonator with gain. 
The corresponding Bloch Hamiltonian is asymmetric~\cite{longhi2015},
\be 
\hat{H} = \left( \begin{array}{cc} 0 & c e^{-\gamma} + e^{-i \bs{k}\cdot \bs{a}_1 } + e^{-i \bs{k}\cdot \bs{a}_2} \\ c e^{\gamma} + e^{i \bs{k}\cdot \bs{a}_1 } + e^{i \bs{k}\cdot \bs{a}_2} & 0 \end{array}\right), 
\label{eq:example_1}
\ee
where $c$ is the intracell coupling strength, $\gamma$ is the strength of the gain and loss, and the other two Hermitian couplings have been normalized to 1. Here, $\bs{k}$ is the Bloch wavevector and the lattice vectors are $\bs{a}_1 = \frac{a}{2} (1, \sqrt{3})$, $\bs{a}_2 = \frac{a}{2}(-1,\sqrt{3})$, $\bs{a}_3 = a(1,0) = \bs{a}_1 - \bs{a}_2$, where $a$ is the lattice period. 

Note that the Hamiltonian (\ref{eq:example_1}) has a chiral symmetry: $\{ \hat{H}, \hat{\sigma}_z \} = 0$. For the bulk-edge correspondence to hold in a lattice, parity-time ($\mathcal{PT}$) symmetry was also required in \cite{Lee}. Here $\mathcal{P} = (y \to -y) \bigotimes \hat{\sigma}_x$ is the reflection $y \to -y$ (which swaps the two sulattices), and $\mathcal{T} = \hat{\sigma}_y K$, where $K$ is complex conjugation. The operator $\mathcal{T}$ takes a nontrivial form because Eq.~(\ref{eq:example_1}) is written in the basis of circulating modes, which is not $\mathcal{T}$-symmetric \cite{Schomerus}.

Akin to Eq.~(1), the Hamiltonian (\ref{eq:example_1}) can be parametrized as $\hat{H}={\bm B} \cdot \hat{\bm \sigma}$ (using non-permuted Pauli matrices) with the two-component complex ${\bf B}$-field:
\begin{align}
B_x &=  c \cosh \gamma + 2 \cos \left(\frac{k_x a}{2}\right) \cos \left( \frac{\sqrt{3} k_y a}{2} \right), \nonumber \\
B_y &=  -i c \sinh \gamma + 2 \cos \left(\frac{k_x a}{2}\right) \sin \left( \frac{\sqrt{3} k_y a}{2} \right).
\label{eq:example_1_field}
\end{align}
Notably, if $k_x$ is taken as a fixed parameter, this model becomes equivalent to the 1D model studied in Ref.~\cite{Lee}. 

\begin{figure}
\includegraphics[width=\columnwidth]{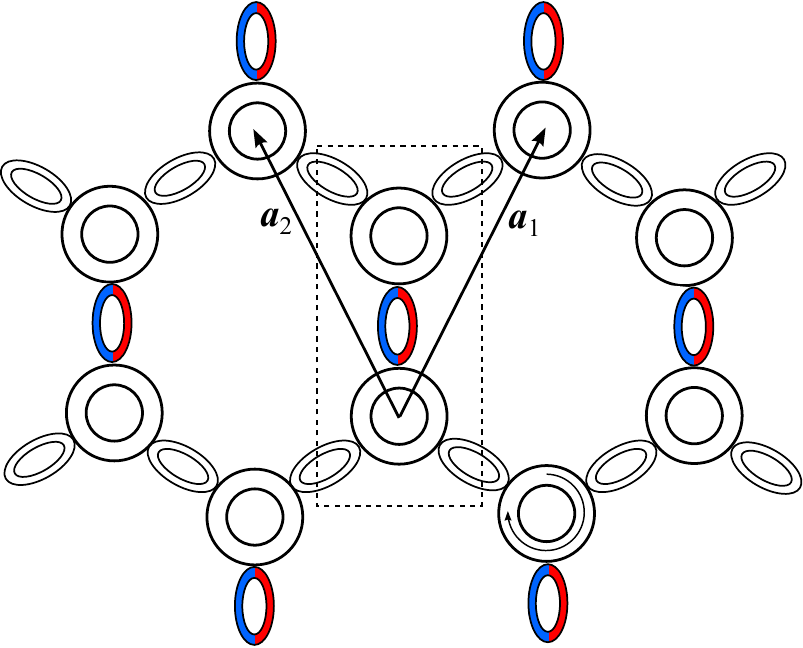}
\caption{Honeycomb lattice formed by ring resonators with gain/loss (red/blue) in one of the three link rings.}
\label{fig:honeycomb_cell}
\end{figure}

\begin{figure}
\includegraphics[width=\columnwidth]{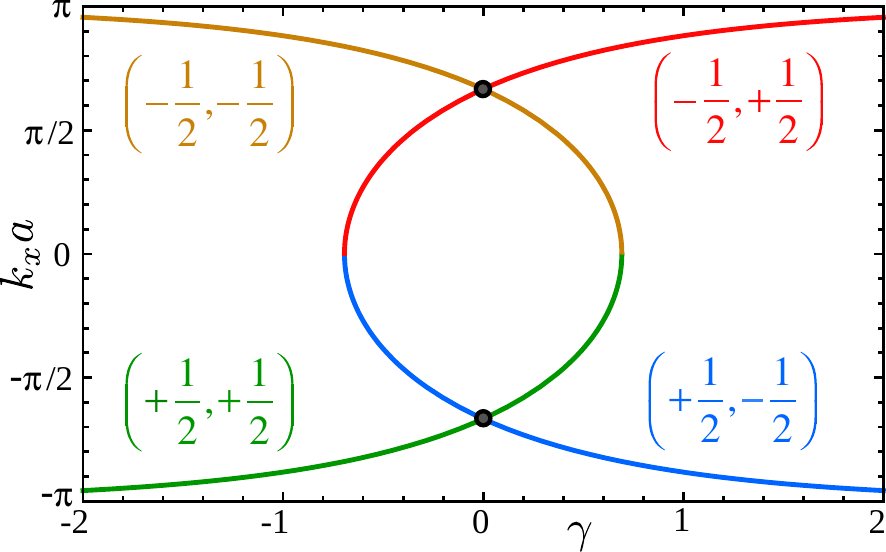}
\caption{Positions of exceptional points with charges ($q_1,q_2$) as a function of the non-Hermitian parameter $\gamma$. Black points indicate the Dirac points in the Hermitian limit $\gamma = 0$.}
\label{fig:honeycomb_EPs}
\end{figure}

In the Hermitian limit $\gamma = 0$, the isotropic ($c=1$) lattice hosts Dirac points (Hermitian degeneracies) at the $\bs{K}$ points, $\bs{K}_{\pm} = \frac{2 \pi}{3a} (\pm 1,\sqrt{3})$. Reciprocal lattice vectors $\bs{G}_1 = \frac{2\pi}{a}(1,\frac{1}{\sqrt{3}})$, $\bs{G}_2 = \frac{2\pi}{a} (-1, \frac{1}{\sqrt{3}})$, $\bs{G}_3 = \frac{2\pi}{a}(1,0)$ relate equivalent $\bs{K}$ points. The anisotropy $c \ne 1$ shifts the Dirac points along the $k_x$ axis (i.e. in the $\bs{G}_3$ direction) towards their partners. They merge and annihilate at the critical points $c=0,2$. A nonzero $\gamma$ splits each Dirac point into a pair of EPs with charges $(q_1,q_2)$ and $k_x$-positions
\begin{align}
\left(+\frac{1}{2},-\frac{1}{2}\right), \quad & k_x = -2\sec^{-1}\! \left(\frac{2 e^{\gamma}}{c}\right) , \nonumber \\
\left(-\frac{1}{2},+\frac{1}{2}\right), \quad & k_x = 2\sec^{-1}\! \left(\frac{2 e^{\gamma}}{c}\right), \nonumber \\
\left(+\frac{1}{2},+\frac{1}{2}\right), \quad & k_x = -2 \cos^{-1}\! \left(\frac{c e^{\gamma}}{2}\right), \nonumber \\
\left(-\frac{1}{2},-\frac{1}{2}\right), \quad & k_x = 2 \cos^{-1}\! \left(\frac{c e^{\gamma}}{2}\right).
\end{align}
Figure~\ref{fig:honeycomb_EPs} plots the EP positions as a function of $\gamma$ (assuming $c=1$). Note the vanishing total charge when EPs coalesce and annihilate at $\gamma = \pm \ln \frac{c}{2}$. On the other hand, when EPs coalesce to form Dirac points, $\sum q_1 = \pm 1$ and $\sum q_2 = 0$. By tuning $(c, \gamma)$ one can control the relative positions and number of EPs in the bulk Hamiltonian. Note that the $m=0$ limit of our continuum Hamiltonian (1) (coalescence of EPs with $\sum q_2 \ne 0$) is not realized in this model. 

\begin{figure}
\includegraphics[width=0.75\columnwidth]{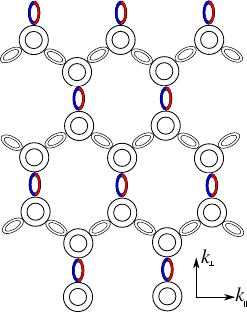}
\caption{Honeycomb lattice with ``bearded'' edge.}
\label{fig:honeycomb_wall}
\end{figure}

We now consider a lattice with ``bearded'' edges shown in Fig.~\ref{fig:honeycomb_wall}. This edge termination respects both the chiral and $\mathcal{PT}$ symmetries. To determine the resulting edge states, one must project $\bs{k} = k_{\parallel} \bs{e}_{\parallel} + k_{\perp} \bs{e}_{\perp}$ onto components parallel and perpendicular to the boundary, described by the reciprocal space basis vectors $\bs{e}_j = \bs{\Gamma}_j/|\bs{\Gamma}_j|$, with $\bs{\Gamma}_{\parallel} = \frac{2\pi}{a} (1,0)$ and $\bs{\Gamma}_{\perp} = \frac{4\pi}{a\sqrt{3}} (0,1)$~\cite{delplace2011}. $k_{\parallel} \in \frac{\pi}{a} [-1,1]$ becomes a parameter and the winding numbers of each domain are calculated over the 1D Brillouin zone defined by $k_{\perp} \in \frac{2\pi}{a\sqrt{3}} (-1,1)$ using Eqs.~(8)~and~(10) for the field (\ref{eq:example_1_field}) (see also Section I above). 

\begin{figure}[h]
\includegraphics[width=\columnwidth]{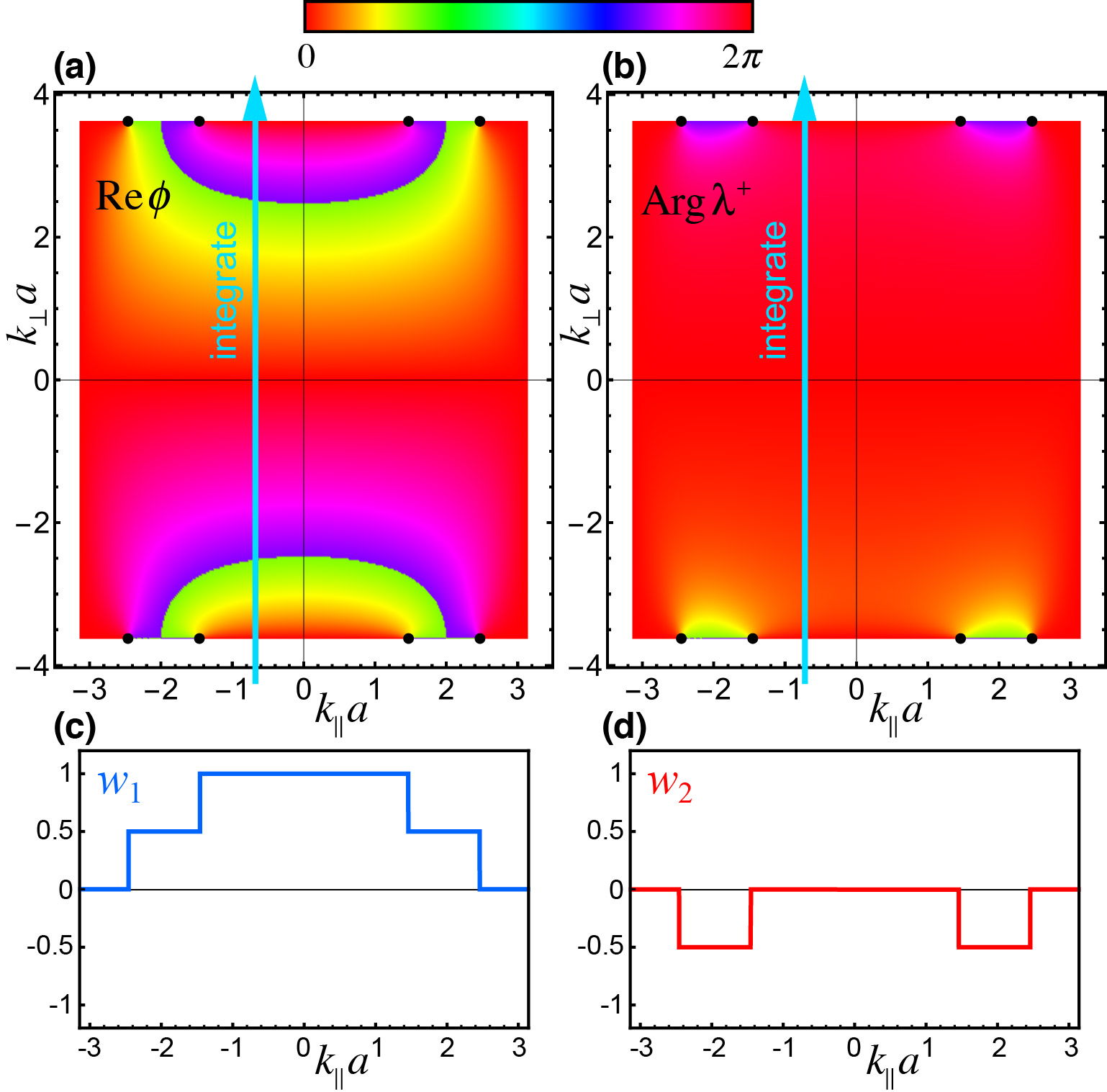}
\caption{Color-coded ${\bf k}$-distributions of (a) the direction ${\rm Re}\,\phi$ (b) the phase ${\rm Arg}\,B = {\rm Arg}\,\lambda^+$ of the complex ${\bf B}$-field (\ref{eq:example_1_field}). Black dots indicate the EPs. Integration of the $k_{\perp}$-gradients of these angles along the $k_{\parallel}$-dependent cyan contour yields the two winding numbers $w_1 (k_y)$ and $w_2 (k_y)$, Eqs.~(8) and (10), shown in panels (c) and (d).
Here parameters are $c=1$ and $\gamma = 0.4$.}
\label{fig:honeycomb_windings}
\end{figure}

Figure~\ref{fig:honeycomb_windings} shows ${\bf k}$-distributions of the direction angle ${\rm Re}\,\phi$ and the phase ${\rm Arg}\,B$ of the ${\bf B}$-field (\ref{eq:example_1_field}), as well as the corresponding winding numbers $w_1(k_{\parallel})$ and $w_2(k_{\parallel})$ for $\gamma = 0.4$. 
One can see that tuning $k_{\parallel}$ through the EPs changes the winding numbers by $\pm \frac{1}{2}$.
Furthermore, there are $k_{\parallel}$-intervals with $w_1 = 0, \frac{1}{2}, 1$ and $w_2 = 0, -\frac{1}{2}\mathrm{sgn}(\gamma)$. Therefore ``bearded'' edges or domain walls between regions with different $\gamma_1 \ne \gamma_2$ can host the non-Hermitian edge modes discussed in the main text. 

To verify the existence of the predicted edge states, we numerically diagonalized $\hat{H}$ on a semi-infinite strip with bearded edges. For sufficiently small $k_x$ both edges support localized modes and they have opposite chiralities. Increasing $k_x$, one of the edge states becomes more strongly localized, while the other becomes more weakly localized, disappearing when the EP is crossed and $w_{1,2}$ become fractional. The sole remaining edge state is defective, i.e. there are two zero energy eigenvalues sharing the same eigenvector. 

This model also provides a simple way to understand the emergence of the anomalous edge states. For $\gamma > 0$, hopping in the $+y$ direction is accompanied by amplification which will counteract the evanescent decay of a state localized to the lower edge, which delocalizes when the amplification rate exceeds the evanescent decay rate and ceases to exist. In contrast, hopping in the $-y$ direction is accompanied by attenuation, which enhances the localization of the zero-energy states at the upper edge.

\subsection{B. Model 2}

\begin{figure}[h]
\includegraphics[width=\columnwidth]{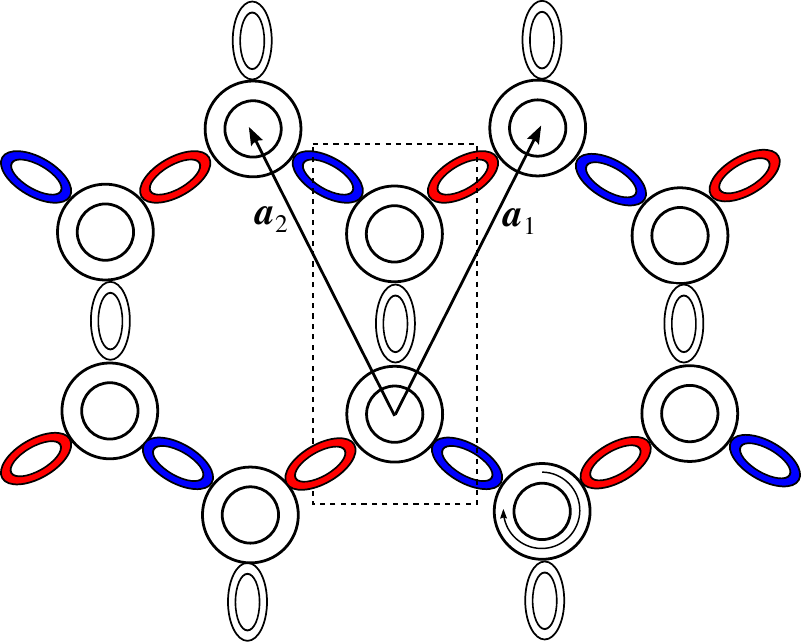}
\caption{Honeycomb lattice formed by ring resonators with dissipative coupling (gain/loss in two of the three link rings).}
\label{fig:honeycomb_cell_2}
\end{figure}

To realize a lattice counterpart of our continuum Hamiltonian Eq.~(1), we require a non-Hermitian term that is asymmetric in $k$. This can be achieved if a pair of links are given balanced gain and loss, as shown in Fig.~\ref{fig:honeycomb_cell_2} and described by the Bloch Hamiltonian
\begin{widetext}
\be 
\hat{H} = \left( \begin{array}{cc} 0 & c + (1-i\gamma) e^{-i \bs{k}\cdot \bs{a}_1 } + (1+i\gamma) e^{-i \bs{k}\cdot \bs{a}_2} \\ c + (1-i \gamma) e^{i \bs{k}\cdot \bs{a}_1 } + (1+i\gamma)e^{i \bs{k}\cdot \bs{a}_2} & 0 \end{array}\right).
\label{eq:example_2}
\ee
Here, $\gamma$ is the dissipative coupling strength describing the gain/attenuation imposed by passing through one of the link resonators~\cite{longhi2016}. This model shares the same chiral and $\mathcal{PT}$ symmetries as Eq.~\eqref{eq:example_1}, but now the non-Hermitian term (the imaginary part of the effective ${\bf B}$-field) is asymmetric in $k_x$:
\begin{align}
B_x &=  c + 2 \cos \left(\frac{k_x a}{2}\right) \cos \left( \frac{\sqrt{3} k_y a}{2} \right) + 2i\gamma \sin \left(\frac{k_x a}{2} \right) \sin \left( \frac{\sqrt{3} k_y a}{2} \right), \nonumber \\
B_y &=  2 \cos \left(\frac{k_x a}{2}\right) \sin \left( \frac{\sqrt{3} k_y a}{2} \right) - 2 i \gamma \sin \left( \frac{k_x a}{2} \right) \cos \left( \frac{\sqrt{3} k_y a}{2} \right).
\label{eq:example_2_field}
\end{align}
\end{widetext}
%

\begin{figure}
\includegraphics[width=\columnwidth]{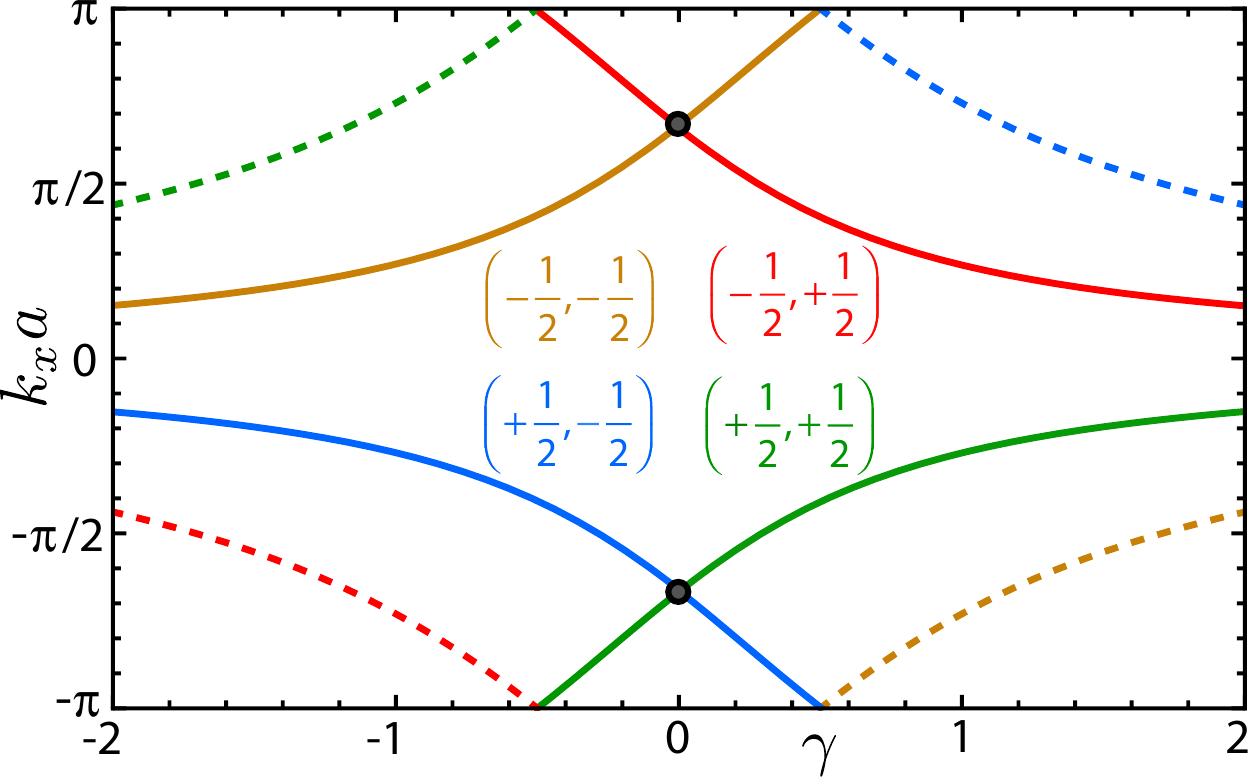}
\caption{Positions of exceptional points with charges ($q_1,q_2$) as a function of the non-Hermitian parameter $\gamma$. Solid lines indicate the EPs at the Brillouin zone edge ($k_y = 2 \pi / \sqrt{3}$), while dashed lines correspond to the zone centre $k_y = 0$. Black points indicate the Dirac points in the Hermitian limit $\gamma = 0$.}
\label{fig:honeycomb_EPs_2}
\end{figure}

Similarly to above, a small $\gamma$ splits the Dirac points confined to the line $k_y a = 2 \pi /\sqrt{3}$ into a pair of EPs with positions $k_x$ plotted in Fig.~\ref{fig:honeycomb_EPs_2}. However, increasing $\gamma$ now leads to the coalescence of an EP pair with the same non-Hermitian charge $q_2$ at the Brillouin zone edge $k_x a = \pm \pi$. Since their total charge is nonzero they cannot annihilate; instead they enter the second Brillouin zone $|k_x a| > \pi$, which is equivalent to the line $k_y = 0$ in the first Brillouin zone (dashed lines in Fig.~\ref{fig:honeycomb_EPs_2}). Meanwhile, the second pair of EPs which approaches $k_x = 0$ reproduces the continuum Hamiltonian Eq.~(1): Expanding $\bs{k} = (0,\frac{2\pi}{\sqrt{3}a}) + \bs{p}$ we obtain, to first order in the displacement $\bs{p}$, the effective ${\bf B}$-field components $B_x \simeq c-2$ (a mass term) and $B_y \simeq i \gamma p_x a + \sqrt{3} p_y a$ (an anisotropic momentum term). 

\begin{figure}
\includegraphics[width=\columnwidth]{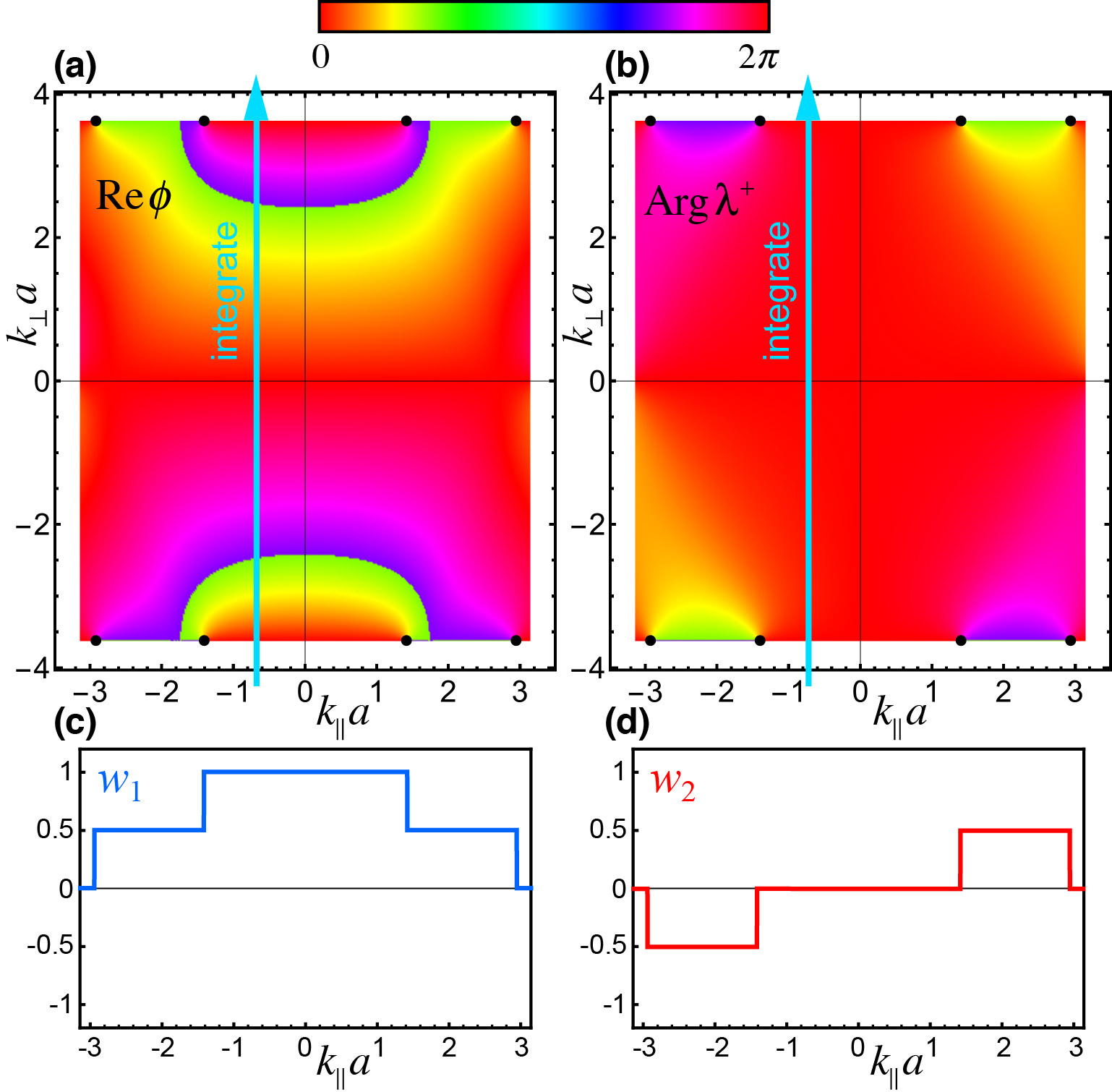}
\caption{Same as in Fig.~\ref{fig:honeycomb_windings} but for the tight-binding model 2 with the Hamiltonian (\ref{eq:example_2}) and ${\bf B}$-field (\ref{eq:example_2_field}).
Parameters are $c=1$ and $\gamma = 0.4$.}
\label{fig:honeycomb_windings_2}
\end{figure}

Similarly to the previous examples, in Fig.~\ref{fig:honeycomb_windings_2} we plot the ${\rm Re}\, \phi ({\bf k})$ and ${\rm Arg}\, \lambda^{+} ({\bf k})$ distributions, together with the corresponding winding numbers $w_1(k_{\parallel})$ and $w_2(k_{\parallel})$, for the Hamiltonian (\ref{eq:example_2}) and ${\bf B}$-field (\ref{eq:example_2_field}). The resulting winding number for a bearded edge $w_1$ ($w_2$) is symmetric (antisymmetric) in $k_{\parallel}$, which provides a tight-binding-model analogue of Eqs.~(9) and (11) and Fig.~(S1).

Finally, let us remark on the key difference between the two lattice models considered here and previously-studied models of honeycomb lattices with $\mathcal{PT}$ symmetry based on a non-Hermitian sublattice potential $B_z = i \gamma \hat{\sigma}_z \ne 0$~\cite{Szameit,Ramezani,Bagarello2016}. In addition to reflection symmetry $\mathcal{P}$, two-dimensional lattices can have rotational symmetry $\mathcal{R}$. In this case (occurring in Ref.~\cite{Szameit,Ramezani,Bagarello2016}), the Bloch Hamiltonian $\hat{H}(\bs{k})$ has a $\mathcal{PT}$ symmetry at \emph{every} wavevector $\bs{k}$ and the energy eigenvalues are always purely real or imaginary. This constraint prevents the appearance of isolated exceptional points, which require complex eigenvalues, and instead results in the splitting of Hermitian Dirac points into ring degeneracies, which are unstable under further perturbations that break the $\mathcal{PT}$ symmetry. In contrast, in our models Eq.~\eqref{eq:example_1} and Eq.~\eqref{eq:example_2} the $\mathcal{PT}$ symmetry relates the Bloch Hamiltonians at inequivalent momenta $\hat{H}(k_x,k_y)$ and $\hat{H}(-k_x,k_y)$. Therefore the Bloch wave spectrum is in general complex, allowing the appearance of isolated non-Hermitian degeneracies (that are stable against small perturbations) and nonzero winding numbers $w_2$.

\section{III. Counting zero modes with the index theorem}

Given a non-Hermitian Hamiltonian $\hat{H}$, we can define the Hermitian Hamiltonian
\begin{equation}
  \hat{\mathcal{H}} = \hat{H}^\dagger \hat{H}.
\end{equation}
If $\hat{H}|\psi\rangle = 0$, then $\hat{\mathcal{H}}|\psi\rangle = 0$.  Conversely, if $\hat{\mathcal{H}}|\psi\rangle = 0$, then $\langle\psi|\hat{H}^\dagger\hat{H}|\psi\rangle = 0$, which implies that $\hat{H}|\psi\rangle = 0$.  Note that this holds for any choice of inner product, and does not rely on the eigenvectors of $\hat{H}$ being orthogonal (they generally are not). Using the non-Hermitian Hamiltonian $\hat{H}$ from Eq.~(5), we obtain
\begin{equation}
  \hat{\mathcal{H}} = (p_{x}-\hat{\sigma}_y sA_{x})^{2}+(sp_{y}-\hat{\sigma}_y A_{y})^{2}+\hat{\sigma}_y B(x,y)
\end{equation}
where $p_i = -i \partial_i$, $\mathbf{A} = (s\mathrm{Im}(m), \mathrm{Re}(m))$, and $B(x,y)=\partial_{x}A_{y}-\partial_{y}A_{x}$.

Let us project to the eigenspace of $\hat{\sigma}_y$, thus replacing $\hat{\sigma}_y$ with $\sigma = \pm 1$.  Now the zero modes of $\hat{\mathcal{H}}$ can be counted via a procedure originally introduced by Aharonov and Casher \cite{AharonovCasher}.  Define the canonical momentum operators
\begin{equation}
\pi_{x}=p_{x}-s \sigma A_{x}\;;\;\pi_{y}=s p_{y}-\sigma A_{y},
\end{equation}
which obey the commutation relations $[\pi_{x},\pi_{y}] = i\sigma B$. The Hermitian Hamiltonian $\hat{\mathcal{H}}$ can then be written as
\begin{equation}
\hat{\mathcal{H}}=(\pi_{x}+i\pi_{y})(\pi_{x}-i\pi_{y}).
\end{equation}
As we have argued above, a zero mode $|\psi\rangle$ must satisfy $(\pi_{x}-i\pi_{y})|\psi\rangle = 0$.  In terms of the wavefunction,
\begin{equation}
  (-i\partial_{x}-s\partial_{y}-s\sigma A_{x}+i\sigma A_{y})\psi(x,y)
  = 0. \label{eq:zero}
\end{equation}
Assuming $\nabla \cdot \bs{A} = 0$, we can let
\begin{equation}
  A_{x}=\partial_{y}\varphi\;,\; A_{y}=-\partial_{x}\varphi. \label{eq:complex_analytic}
\end{equation}
With this gauge choice, the magnetic field is the source of a potential $\varphi$:
\begin{equation}
\nabla^{2}\varphi(x,y)=B(x,y).
\label{eq:green}
\end{equation}
We substitute this back into Eq.~(\ref{eq:zero}), and make the further gauge substitution
\begin{equation}
\psi(x,y)=\exp[-\sigma\varphi(x,y)]f(x,y).
\label{fdef}
\end{equation}
Then $f(x,y)$ obeys
\begin{equation}
(\partial_{x}-is \partial_{y})f(x,y)=0.
\end{equation}

In the first case of interest, $s_1=s_2=1$, $f(x,y)$ is analytic in the complex plane, and hence cannot be normalized. Thus, the normalization of $\psi$ must arise from the exponential factor in Eq.~(\ref{fdef}). Using Eq.~(\ref{eq:green}), we write
\begin{align}
  \varphi(r)&=\int dr'G(r-r')B(r'),\\
  G(r-r')&=\frac{1}{2\pi}\ln\!\left(\frac{|r-r'|}{l_{B}}\right).
\end{align}
Here, $l_{B}$ is the magnetic length that serves as the cut-off of the theory. For $r\gg r'$,
\begin{equation}
\varphi(r) \rightarrow \mbox{ln}\!\left(\frac{r}{l_{B}}\right)^{\!\Phi/2\pi},
\end{equation}
where
\begin{equation}
\Phi=\iint dx\, dy\,(\partial_{x}A_{y}-\partial_{y}A_{x})=2\pi(N+\epsilon).
\end{equation}
Here, $N$ is an integer and $0<\epsilon<1$.  Substituting this back into Eq.~(\ref{fdef}), we arrive at
\begin{equation}
\psi(x,y)=\left(\frac{r}{l_{B}}\right)^{\!-\sigma{\Phi}/{2\pi}}\!f(x,y).
\end{equation}
Next, we can expand the analytic function $f(x,y)$ with
\begin{equation}
f(x,y)=f(z)=z^{j},
\end{equation}
where $z=x+iy$.  The resulting wave function is
\begin{equation}
\psi_{j}(z)=\left(\frac{|z|}{l_{B}}\right)^{\!-\sigma\Phi/2\pi}\!z^{j}.
\end{equation}
For $\sigma\Phi>0$, we require $j=0,1,2,\dots,N-1$, so that $\psi_{j}(z)$ is normalizable.  For $\sigma\Phi<0$, however, there are no normalizable zero modes.  Thus, $\hat{\mathcal{H}}$ has a total of $N$ zero modes.  The zero modes are polarized either spin-up ($\sigma=+1$) or spin-down ($\sigma=-1$), depending on the sign of the total magnetic flux $\Phi$. The flux $\Phi$ is determined by the mass profile $m(x,y) = m(z)$, which must be a complex analytic function according to Eq.~\eqref{eq:complex_analytic}.

In the second case, $s_1 = -s_2 = s$, $f(x,y)= f(z)$ is a piecewise-analytic function which can be decomposed in medium 1 using the Cauchy integral formula as
\be 
f(z) = \frac{1}{2\pi i} \oint \frac{f(t)}{t-z} dt,
\ee
where the integral is over the boundary between the two media. The analytic function $f^*(z)$ in medium 2 can be obtained similarly by requiring the continuity of $\psi$ at the interface. Since $f(z)$ is now normalizable, zero modes can exist even when the flux $\Phi$ vanishes, in which case they will be spin-degenerate. A nonzero flux $\Phi$ will generate an imbalance between the number of spin-up and spin-down modes.

Suppose now that $m=m(x)$ forms a straight domain wall in the vicinity of $x=0$ with mass parameters $m_{1,2}$ in the limit $|x| \rightarrow \infty$. A straightforward calculation shows that $\varphi(x) \rightarrow -m_1 x$, $\varphi(x\rightarrow \infty) = -m_2 x$, the magnetic flux per unit length is $(m_2 - m_1)$, and $f(x,y) \propto \exp({-s_{1,2} k x + i k y})$. The resulting wavefunction is
\be 	
\psi(x,y) \sim \exp\left[{i k y - s_{1,2} k x - \sigma\varphi(x)}\right],
\ee
reproducing the conditions for normalizable zero modes, Eqs.~(7), discussed in the main text.


\vspace*{-0.2cm}

\begin{thebibliography}{99}


\bibitem{Thouless} D.~J.~Thouless, {\it Topological Quantum Numbers in Nonrelativistic Physics} (World Scientific, 1998).

\bibitem{TI_reviews}
M.~Z. Hasan and C.~L. Kane, \textit{Colloquium: Topological insulators}, Rev.~Mod.~Phys.~{\bf 82}, 3045 (2010); 
X.-L. Qi and S.-C. Zhang, \textit{Topological insulators and superconductors}, Rev.~Mod.~Phys.~{\bf 83}, 1057 (2011).

\bibitem{Lu2014}
L.~Lu, J.D.~Joannopoulos, and M.~Solja\u{c}i\'{c}, \textit{Topological photonics}, Nat.~Photon.~{\bf 8}, 821 (2014).

\bibitem{non-Hermitian_review}
N.~Moiseyev, {\it Non-Hermitian Quantum Mechanics} (Cambridge Univ. Press, 2011).

\bibitem{Bender2007}
C.~M. Bender, \textit{Making sense of non-Hermitian Hamiltonians}, Rep.~Prog.~Phys.~{\bf 70}, 947 (2007).

\bibitem{Cao2015}
H.~Cao and J.~Wiersig, \textit{Dielectric microcavities: Model systems for wave chaos and non-Hermitian physics}, Rev.~Mod.~Phys.~{\bf 87}, 61 (2015).

\bibitem{dirac_review_1}
M.~Goerbig and G.~Montambaux, \textit{Dirac fermions in condensed matter and beyond}, Seminaire Poincare {\bf 17}, 1 (2013) (arXiv:1410.4098).

\bibitem{dirac_review_2}
D.~Leykam and A.~S.~Desyatnikov, \textit{Conical intersections for light and matter waves}, Advances in Physics: X {\bf 1}, 101 (2016). 

\bibitem{Berry2004}
M.V. Berry, \textit{Physics of nonhermitian degeneracies}, Czech.~J.~Phys.~{\bf 54}, 1039 (2004).

\bibitem{Heiss2012} W.~D.~Heiss, \textit{The physics of exceptional points}, J.~Phys.~A: Math.~Theor.~{\bf 45}, 444016 (2012).

\bibitem{Rudner2009} M.~S. Rudner and L.~S. Levitov, \textit{Topological transition in a non-Hermitian quantum walk}, Phys.~Rev.~Lett.~{\bf 102}, 065703 (2009).

\bibitem{Esaki} K.~Esaki, M.~Sato, K.~Hasebe, and M.~Kohmoto, \textit{Edge states and topological phases in non-Hermitian systems}, Phys.~Rev.~B {\bf 84}, 205128 (2011).

\bibitem{hu2011} Y.~C. Hu and T.~L. Hughes, \textit{Absence of topological insulator phases in non-Hermitian PT-symmetric Hamiltonians}, Phys.~Rev.~B {\bf 84}, 153101 (2011).

\bibitem{Schomerus} H. Schomerus, \textit{Topologically protected midgap states in complex photonic lattices}, Opt. Lett. {\bf 38}, 1912 (2013).

\bibitem{Malzard2015}
S.~Malzard, C.~Poli, and H.~Schomerus, \textit{Topologically protected defect states in open photonic systems with non-Hermitian charge-conjugation and parity-time symmetry}, Phys.~Rev.~Lett.~{\bf 115}, 200402 (2015).

\bibitem{Lee} T.~E.~Lee, \textit{Anomalous edge state in a non-Hermitian lattice}, Phys.~Rev.~Lett.~{\bf 116}, 133903 (2016).

\bibitem{Diehl2011} S.~Diehl {\it et al.}, \textit{Topology by dissipation in atomic quantum wires}, Nat.~Phys.~{\bf 7}, 971 (2011).

\bibitem{Zeuner2015}
J.~M.~Zeuner {\it et al.}, \textit{Observation of a topological transition in the bulk of a non-Hermitian system}, Phys.~Rev.~Lett.~{\bf 115}, 040402 (2015).

\bibitem{Yuce2015}
C.~Yuce, \textit{Topological phase in a non-Hermitian $PT$ symmetric system}, Phys.~Lett.~A {\bf 379}, 1213 (2015).

\bibitem{SanJose2016}
P.~San-Jose {\it et al.}, \textit{Majorana bound states from exceptional points in non-topological superconductors}, Sci.~Rep.~{\bf 6}, 21427 (2016).

\bibitem{Gonzales2016}
J. Gonz\'{a}les and R.A. Molina, \textit{Macroscopic degeneracy of zero-mode rotating surface states in 3D Dirac and Weyl semimetals under radiation}, Phys.~Rev.~Lett.~{\bf 116}, 156803 (2016).

\bibitem{Rudner2016}
M.~S.~Rudner, M.~Levin, and L.~S.~Levitov, \textit{Survival, decay, and topological protection in non-Hermitian quantum transport}, arXiv:1605.07652 (2016).

\bibitem{JR} R.~Jackiw and C.Rebbi, \textit{Solitons with fermion number 1/2}, Phys.~Rev.~D {\bf 13}, 3398 (1976).

\bibitem{Shen2011}
S.-Q.~Shen, W.-Y.~Shan, and H.-Z.~Lu, \textit{Topological insulator and the Dirac equation}, SPIN {\bf 1}, 1 (2011).

\bibitem{Schnyder} A.~P.~Schnyder, S.~Ryu, A.~Furusaki, and A.~W.~W.~Ludwig, \textit{Classification of topological insulators and superconductors in three spatial dimensions}, Phys.~Rev.~B {\bf 78}, 195125 (2008).

\bibitem{SSH} W.~P.~Su, J.~R.~Schrieffer, and A.~J.~Heeger, \textit{Solitons in polyacetylene}, Phys.~Rev.~Lett.~{\bf 42}, 1698 (1979).

\bibitem{Kitaev}
A.Y. Kitaev, \textit{Unpaired Majorana fermions in quantum wires}, Phys.-Uspekhi {\bf 44}, 131 (2001).

\bibitem{AharonovCasher} Y.~Aharonov and A.~Casher, \textit{Ground state of a spin-$^1\!/_2$ charged particle in a two-dimensional magnetic field}, Phys.~Rev.~A {\bf 19}, 2461 (1979).

\bibitem{longhi2015}
S.~Longhi, D.~Gatti, and G.~Della Valle, \textit{Non-Hermitian transparency and one-way transport in low-dimensional lattices by an imaginary gauge field}, Phys. Rev. B {\bf 92}, 094204 (2015).

\bibitem{supplemental}
See Supplemental Material, which includes Refs.~\cite{delplace2011,longhi2016,Bagarello2016}.

\bibitem{delplace2011} P. Delplace, D. Ullmo, and G. Montambaux, \textit{Zak phase and the existence of edge states in graphene}, Phys. Rev. B {\bf 84}, 195452 (2011).

\bibitem{longhi2016}
S. Longhi, \textit{Non-Hermitian tight-binding network engineering}, Phys. Rev. A {\bf 93}, 022102 (2016).

\bibitem{Bagarello2016}
F. Bagarello and N. Hatano, \textit{PT-symmetric graphene under a magnetic field}, Proc. R. Soc. A {\bf 472}, 0365 (2016).

\bibitem{doubling}
Similar to the doubling of Hermitian Dirac points required by the Nielsen-Ninomiya theorem, a lattice regularization doubles the degeneracies and introduces a large-wavenumber cutoff to the edge modes; see \cite{supplemental}.

\bibitem{Heiss2001} W. D. Heiss and H. L. Harney, \textit{The chirality of exceptional points}, Eur.~Phys.~J.~D {\bf 17}, 149 (2001).

\bibitem{Dembowski2003}
C.~Dembowski {\it et al.}, \textit{Observation of a chiral state in a microwave cavity}, Phys.~Rev.~Lett.~{\bf 90}, 034101 (2003).

\bibitem{Peng2016}
B.~Peng {\it et al.}, \textit{Chiral modes and directional lasing at exceptional points}, Proc.~Nat.~Acad.~Sci. {\bf 113}, 6845 (2016).

\bibitem{mailybaev2005} A. A. Mailybaev, O. N. Kirillov, and A. P. Seyranian, \textit{Geometric phase around exceptional points}, Phys. Rev. A {\bf 72}, 014104 (2005).

\bibitem{demowski2004}
C.~Dembowski {\it et al.}, \textit{Encircling an exceptional point}, Phys. Rev. E {\bf 69}, 056216 (2004).

\bibitem{gao2015}
T.~Gao {\it et al.}, \textit{Observation of non-Hermitian degeneracies in a chaotic exciton-polariton billiard}, Nature {\bf 526}, 554 (2015).

\bibitem{garrison1988} J. C. Garrison and E. M. Wright, \textit{Complex geometrical phases for dissipative systems}, Phys.~Lett.~A {\bf 128}, 177 (1988).

\bibitem{dennis_review} M.~R.~Dennis, K.~O'Holleran, and M.~J.~Padgett, \textit{Singular optics: optical vortices and polarization singularities}, Prog. Opt. {\bf 53}, 293 (2007).

\bibitem{Jackiw} R.~Jackiw, \textit{Fractional charge and zero modes for planar systems in a magnetic field}, Phys.~Rev.~D {\bf 29}, 2375 (1986).

\bibitem{Szameit} A.~Szameit, M.~C.~Rechtsman, O.~Bahat-Treidel, and M.~Segev, \textit{$PT$-symmetry in honeycomb photonic lattices}, Phys.~Rev.~A \textbf{84}, 021806(R) (2011).

\bibitem{Ramezani} H.~Ramezani, T.~Kottos, V.~Kovanis, and D.~N.~Christodoulides, \textit{Exceptional-point dynamics in photonic honeycomb lattices with $PT$ symmetry}, Phys.~Rev.~A \textbf{85}, 013818 (2012).

\bibitem{Liang2013}
G.~Q.~Liang and Y.~D.~Chong, \textit{Optical resonator analog of a two-dimensional topological insulator},
Phys.~Rev.~Lett.~\textbf{110}, 203904 (2013).

\bibitem{Chong2015}
Y.~D.~Chong and M.~C.~Rechtsman, \textit{Tachyonic dispersion in coherent networks}, J.~Opt.~\textbf{18}, 014001 (2016).

\end{thebibliography}
\end{document}